\documentclass[prd,twocolumn,floatfix,nofootinbib,amsmath,amssymb,floatfix]{revtex4}
\usepackage{graphicx,color,dcolumn,booktabs,bm,mathrsfs}
\usepackage{longtable,lscape}
\usepackage{txfonts}
\usepackage{overpic}
\usepackage{amssymb}
\usepackage{indentfirst}
\usepackage{float}
\usepackage{feynmf}   
\usepackage{slashed}  
\usepackage{cases}
\usepackage{color}
\usepackage{multirow}
\usepackage{epstopdf}
\usepackage{CJK}
\usepackage[colorlinks,
            citecolor=blue,
            anchorcolor=red,
            menucolor=red,
            linkcolor=red,
            filecolor=red,
            runcolor=red,
            urlcolor=blue,
            frenchlinks=true]{hyperref}
\usepackage{subfigure}

\begin{document}
\title{Production of $d_{N\Omega}$ dibaryon in kaon induced reactions}

\author{Jing Liu$^{1}$}\email{jingliu@seu.edu.cn}
\author{Qi-Fang L\"u$^{2,3,4}$} \email{lvqifang@hunnu.edu.cn} %
\author{Chun-Hua Liu$^{1}$}\email{liuch@seu.edu.cn}
\author{Dian-Yong Chen$^{1,5}$\footnote{Corresponding author}}\email{chendy@seu.edu.cn}
\author{Yu-Bing Dong$^{6,7}$} \email{dongyb@ihep.ac.cn}

\affiliation{
$^1$School of Physics, Southeast University, Nanjing 210094, China\\
$^2$Department of Physics, Hunan Normal University, and Key Laboratory of Low-Dimensional Quantum Structures and Quantum Control of Ministry of Education, Changsha 410081, China\\
$^3$Synergetic Innovation Center for Quantum Effects and
Applications (SICQEA), Hunan Normal University, Changsha 410081, China\\
$^4$Research Center for Nuclear Physics (RCNP), Ibaraki,
Osaka 567-0047, Japan\\
$^5$Lanzhou Center for Theoretical Physics, Lanzhou University, Lanzhou 730000, China\\
$^6$Institute of High Energy Physics, Chinese Academy
of Sciences, Beijing 100049, China\\
$^7$School of Physical Sciences, University of Chinese Academy of Sciences, Beijing 101408, China
}
\date{\today}
\begin{abstract}
In this work, we propose to investigate the $d_{N\Omega}$ dibaryon production in the process $K^- p \rightarrow  d_{N\Omega} \bar{\Xi}^0$ by utilizing the kaon beam with the typical momentum to be around 10 GeV, which may be available at COMPASS, OKA@U-70 and SPS@CERN. The cross sections for $K^- p \rightarrow  d_{N\Omega} \bar{\Xi}^0$ are estimated and in particular, the magnitude of the cross sections are evaluated to be several hundreds nanobarn at $P_K=20$ GeV. Considering that $d_{N\Omega}$ dibaryon dominantly decay into $\Xi \Lambda$ and $\Xi \Sigma$, we also estimate the cross sections for $K^- p \to \Xi^0 \Lambda \bar{\Xi}^0$ and $K^- p \to \Xi^- \Sigma^+ \bar{\Xi}^0$, where the $d_{N\Omega}$ dibaryon can be observed in the invariant mass distributions of $\Xi^0 \Lambda$ and $\Xi^- \Sigma^+ $, respectively.
\end{abstract}
\pacs{13.25.GV, 13.75.Lb, 14.40.Pq}

\maketitle

\section{Introduction}
\label{sec:Introduction}

In the strict sense, dibaryon is a sort of particle with baryon number $B=2$, which is composed of six quarks , typically two baryons. As a typical dibaryon, a deuteron consists of one proton and one neutron with a binding energy to be $2.22$ MeV. Besides the deuteron, searching for the dibaryon states composed of other baryons becomes one of intriguing topics of hadron physics in recent decades with the developments of experimental techniques and the accumulations of experimental data.

The theoretical investigations of the dibaryon could date back to the year of 1964, when the dibaryon states were studied by Dyson and Xuong ~\cite{Dyson:1964xwa} with SU(6) theory, and then in 1977 Jaffe predicted the existence of the $H$ and $H^*$ particles with strangeness $S= -2$, which could be the bound state of $\Lambda \Lambda$ and $\Sigma \Sigma$, respectively. The investigations in two different quark models indicated that there should exist dibaryon states with strangeness to be $-3$~\cite{Goldman:1987ma}, which were stable for strong decay ~\cite{Goldman:1987ma}. And later, the mass spectrum of the low-lying dibaryons with strangeness to be -1 were evaluated in the quark model~\cite{Konno:1988nb}. Furthermore, some other models had been extended to study the dibaryons, for examples, the quark-cluster model~\cite{Oka:1988yq}, Skyrme Model~\cite{Kopeliovich:1992sa}, quark potential model ~\cite{Wang:1995bg}, the chiral SU(3) quark model ~\cite{Zhang:2000sv}, quark delocalization and color screening model~\cite{Wu:1996fm,Wang:1992wi, Chen:2007qn,Ping:2000cb,Pang:2003ty} and realistic phenomenological nucleon-nucleon interaction models ~\cite{Julia-Diaz:2004ict, Froemel:2004ea}. In these model, a series of dibaryon states have been investigated, such as the nonstrange dibaryon $d^{\ast}$ with $I(J^P)=0(3^+)$, dibaryons composed of $N\Xi^\prime$, $N\Xi_{c}$, $N\Xi_{cc}$, $\Xi_{cc} \Xi_{cc}$~\cite{Julia-Diaz:2004ict, Froemel:2004ea}, $N\Omega$  and $\Delta \Omega$~\cite{Dai:2006dgq}.

On the experimental side, the first breakthrough was the observations of $d^\ast(2380)$, which was firstly observed in the cross sections for $np \to d \pi \pi$ by CELSIUS/WASA Collaboration in 2009 ~\cite{Bashkanov:2008ih}. The discovery of $d^\ast(2380)$ by CELSIUS/WASA Collaboration has encouraged the experimentists to search $d^\ast(2380)$ in more processes. With much better statistics and precision data sample, the WASA-at-COSY Collaboration found that the observed angular distributions for the deuterons and pions of the process $np \to d\pi^0 \pi^0$ in the center-of-mass system clearly preferred $J=3$~\cite{WASA-at-COSY:2011bjg, WASA-at-COSY:2012seb} for $d^\ast(2380)$, and then the $I(J^{P})$ quantum numbers were determined to be $0(3^+)$. Later on, the properties of $d^\ast(2380)$ had been investigated in various unpolarized $np$ collision processes with more precise data samples, for example,  $np\rightarrow np \pi^0\pi^0$, $np\rightarrow pp\pi^0\pi^-$, $np\rightarrow d \pi^0\pi^0$~\cite{WASA-at-COSY:2014qkg, WASA-at-COSY:2014squ, WASA-at-COSY:2013fzt}, and polarized $np$ scattering process~\cite{WASA-at-COSY:2014dmv}.

The experimental observations of $d^\ast(2380)$ has attracted a lot of theorists' attentions to explore its properties~\cite{Shi:2019dpo, Huang:2014kja,Dong:2015cxa, Dong:2016rva, Dong:2017geu, Huang:2015nja, Gongyo:2020pyy, Huang:2022qgx}. There are also dissenting voices, for example, in Refs.~\cite{Ikeno:2021frl,Molina:2021bwp}, the authors claimed that the peak structure should be tied to a triangle singularity in the last step of the reaction. Anyhow, the experimental breakthrough has stimulate theorists' interests in dibaryons states. Some dibaryon systems had been investigated in various models. For examples, in a three-body hadronic model, the authors in Ref. ~\cite{Gal:2014zia} calculated the $N\Delta$ and $\Delta \Delta$ dibaryon states. In the one pion exchange potential model, the $H$-like $\Lambda_{c}\Lambda_{c}$~\cite{Meguro:2011nr}, $\Lambda_c N $~\cite{Liu:2011xc} dibaryons were researched. The possible $\Lambda_{c}\Lambda_{c}/\Lambda_{b}\Lambda_{b}$~\cite{Huang:2013rla}, and $N\Sigma_{c,b}$~\cite{Huang:2013zva} dibaryon states were studied in the quark delocalization color screening model, and the possible $\Delta^0\Delta^0$, $\Omega\Omega$, $\Xi\Xi$ dibaryons were investigated in Refs.~\cite{Lu:2020qme,Liu:2021gva, Mutuk:2022zgn, Chen:2019vdh, Brambilla:2022ura}.

\begin{figure*}[htb]
\begin{tabular}{ccc}
  \centering
  \includegraphics[width=80mm]{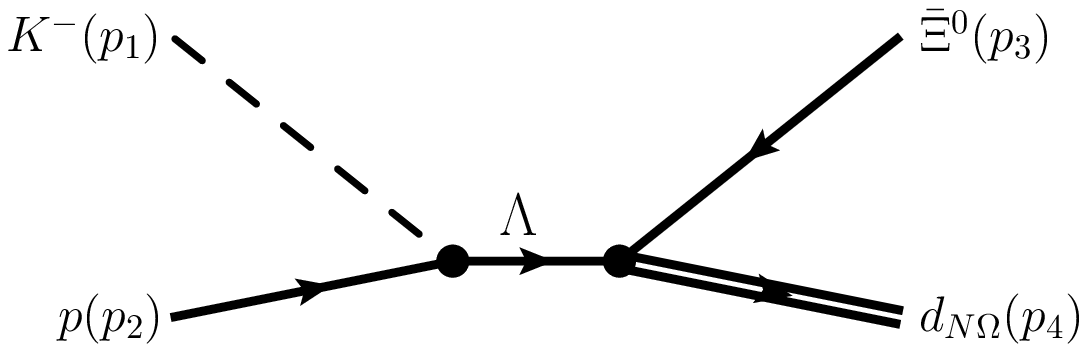}& $\hspace{5mm}$&
  \includegraphics[width=80mm]{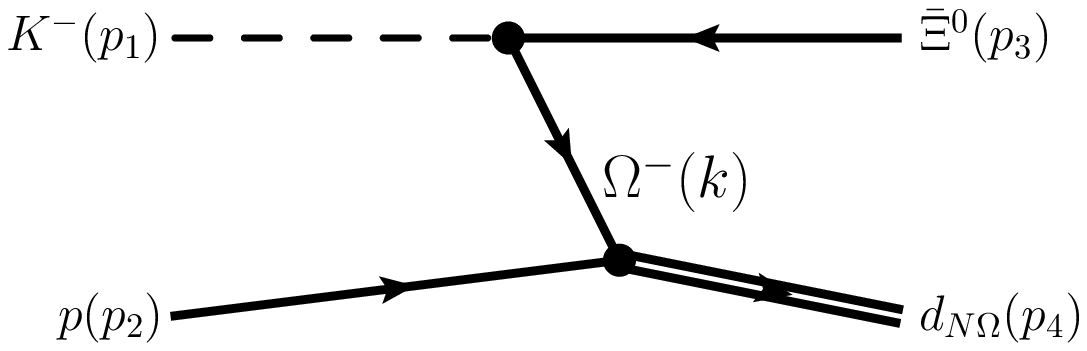}\\
  \\
 {\large(a)} & & {\large(b)} \\
 \end{tabular}
  \caption{Diagrams contributing to the process of $K^-p \to  d_{N\Omega}\bar{\Xi}^0$, where the $d_{N\Omega}$ is considered as a S-wave $N\Omega$ dibaryon with $J^P=2^+$.}\label{Fig:Mech1}
\end{figure*}

As for the dibaryon states with strangeness $S=-3$, the possibilities of the existence of such kind of dibaryons have been evaluated by various model. For example, the estimations in Ref.~\cite{Li:1999bc} indicated that there might exist two bound states  $N\Omega$ and $\Delta\Omega$ dibaryons, respectively, and in Ref.~\cite{Dai:2006dgq}, the authors found that the $N\Omega$ and $\Delta \Omega$ were weakly bound systems in the chiral quark model. The QCD sum rule estimations in Ref.~\cite{Chen:2021hxs} indicate that the $N\Omega$ dibaryon with $J^P=2^+$ was stable for strong decay. In Ref.~\cite{HALQCD:2014okw}, the HAL QCD Collaboration calculated the $N\Omega$ potential in $2+1$ flavor Lattice QCD and one bound state with the binding energy to be about 20 MeV was found. In the year of 2019, the HAL QCD Collaboration updated their estimations of Ref. ~\cite{HALQCD:2014okw} near the physical point and they found the binding energy of the $p\Omega(^5S_2)$ became $2.46(0.34)(^{+0.04}_{-0.11})$ MeV. Simulated by the most recent Lattice QCD estimations, the authors in Ref.~\cite{Zhang:2020dma} estimated the productions of the $\Omega$-dibaryons by utilizing a dynamical coalescence mechanism for the relativistic heavy-ion collisions, and the strong decays of $d_{N\Omega}$ into conventional hadrons were evaluted~\cite{Xiao:2020alj}.

Actually, experimentally producing the $d_{N\Omega}$ dibaryon is the first step of investigating its properties. Thus, searching $d_{N\Omega}$ experimentally becomes a pressing task. The key point of producing $d_{N\Omega}$ dibaryon is the production of $\Omega$ baryon. In the high energy heavy-ion collision or $pp$ collision processes, a large quantity of quarks with different flavors can be produced, where three strange quarks have chance to form a $\Omega$ baryon and produce a $d_{N\Omega}$ dibaryon by interacting with a nucleon. In the year of 2018, the STAR Collaboration at RIHC investigated the proton-$\Omega$ interaction by measuring the corresponding correlation function in heavy-ion collisions at $\sqrt{s_{NN}}=200$ GeV~\cite{STAR:2018uho}. By comparing the measured correlation ratio with the theoretical calculations, they concluded that the measurements slightly favored a proton-$\Omega$ bound system with a binding energy of $~27$ MeV~\cite{STAR:2018uho}. Similarly, the ALICE Collaboration also proposed to investigate the strong interaction among hadrons, including proton-$\Omega$, by using the ultrarelativistic proton-proton collisions at LHC~\cite{ALICE:2020mfd}.

Besides the high energy heavy-ion collision and $pp$ collision processes, the $\Omega$ baryon can also be produced by kaon induced reactions since there is already a strange quark in the kaon. In Ref.~\cite{Aoki:2021cqa}, a project of the extension of the Hadron Experimental Facility at J-PARC was proposed. By utilizing the secondary beam of  kaon with the typical momentum to be around 3 GeV, the $\Omega$ baryon can be produced via the $K^- p \to \Omega^- K^+ \bar{K}^{(\ast)0}$ reaction and then the $d_{N\Omega}$ dibaryon can be investigated in the invariant mass spectrum of $\Xi^- \Lambda$ of the process $\Omega^- d \to \Xi^- \Lambda p$.

In addition to the secondary beam of kaon at J-PARC~\cite{Nagae:2008zz}, the high energy kaon beam with high quality are also available at  COMPASS~\cite{Nerling:2012er}, OKA@U-70~\cite{Obraztsov:2016lhp} and SPS@CERN~\cite{Velghe:2016jjw}. These high energy beam, especially ones with momentum to be around 10 GeV, may provide us another approach to directly produce $d_{N\Omega}$ via the reaction $K^- p \to d_{N\Omega}\bar{\Xi}^0$ and then the $d_{N\Omega}$ dibaryon can decay into $\Xi \Lambda$ and $\Xi \Sigma$. Thus, one can detect $d_{N\Omega}$ in the $\Xi\Lambda$ and $\Xi\Sigma$ invariant mass spectra of the processes $K^- p \to \bar{\Xi}^0 \Xi \Lambda$ and $K^- p \to \bar{\Xi}^0 \Xi \Sigma$, respectively. In the present work, we evaluate the possibility of observing $d_{N\Omega}$ dibaryon in these processes by estimating their cross sections.

This work is organized as follows. After the introduction, the mechanism of $d_{N\Omega}$ production in kaon induced reactions is presented. In Section ~\ref{sec:NR}, the cross sections for the processes $K^- p \rightarrow d_{N\Omega} \bar{\Xi}^0$, $K^- p \rightarrow \Xi^0\Lambda \bar{\Xi}^0$ and $K^- p \rightarrow \Xi^-\Sigma^+\bar{\Xi}^0$ are presented, and the last section is devoted to a short summary.\\

\begin{figure*}[htb]
   \includegraphics[width=75mm]{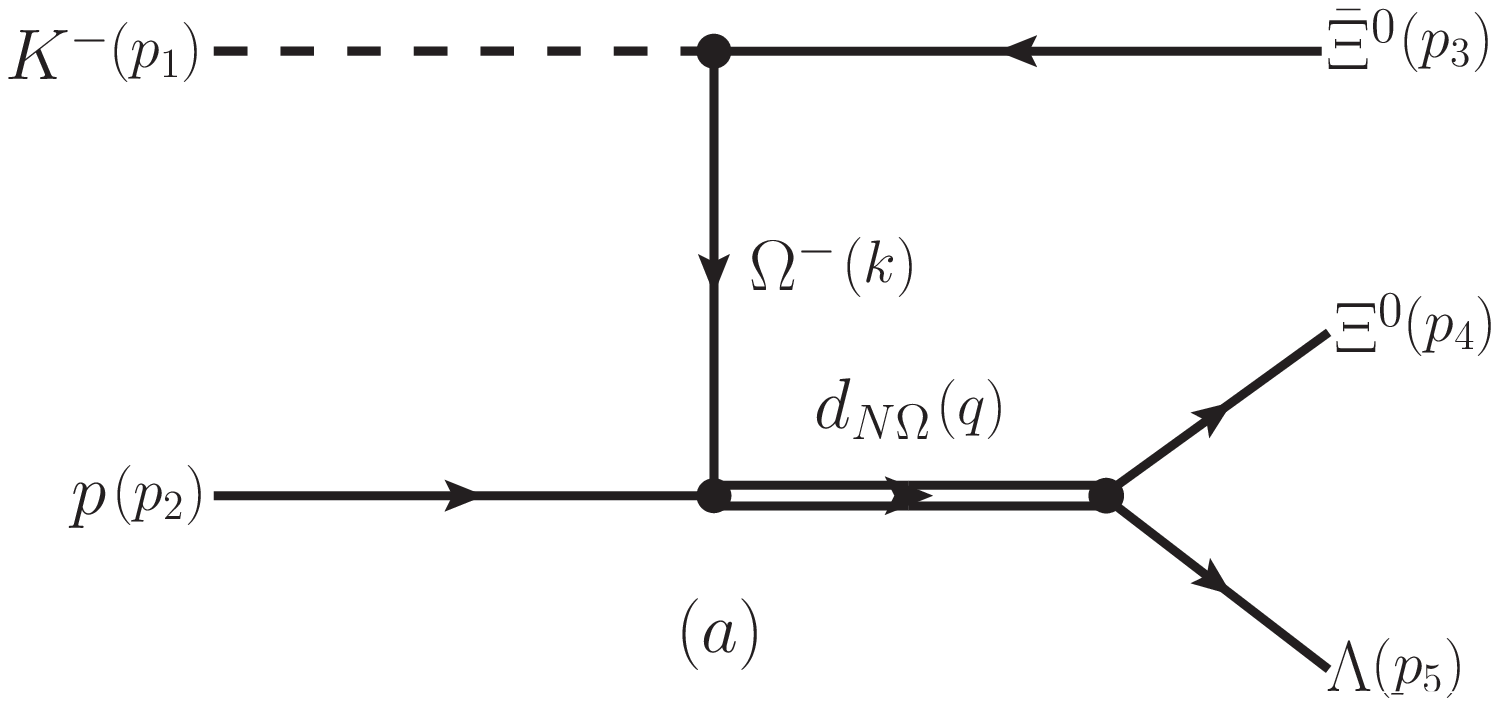}
   \hspace{10mm}
  \includegraphics[width=75mm]{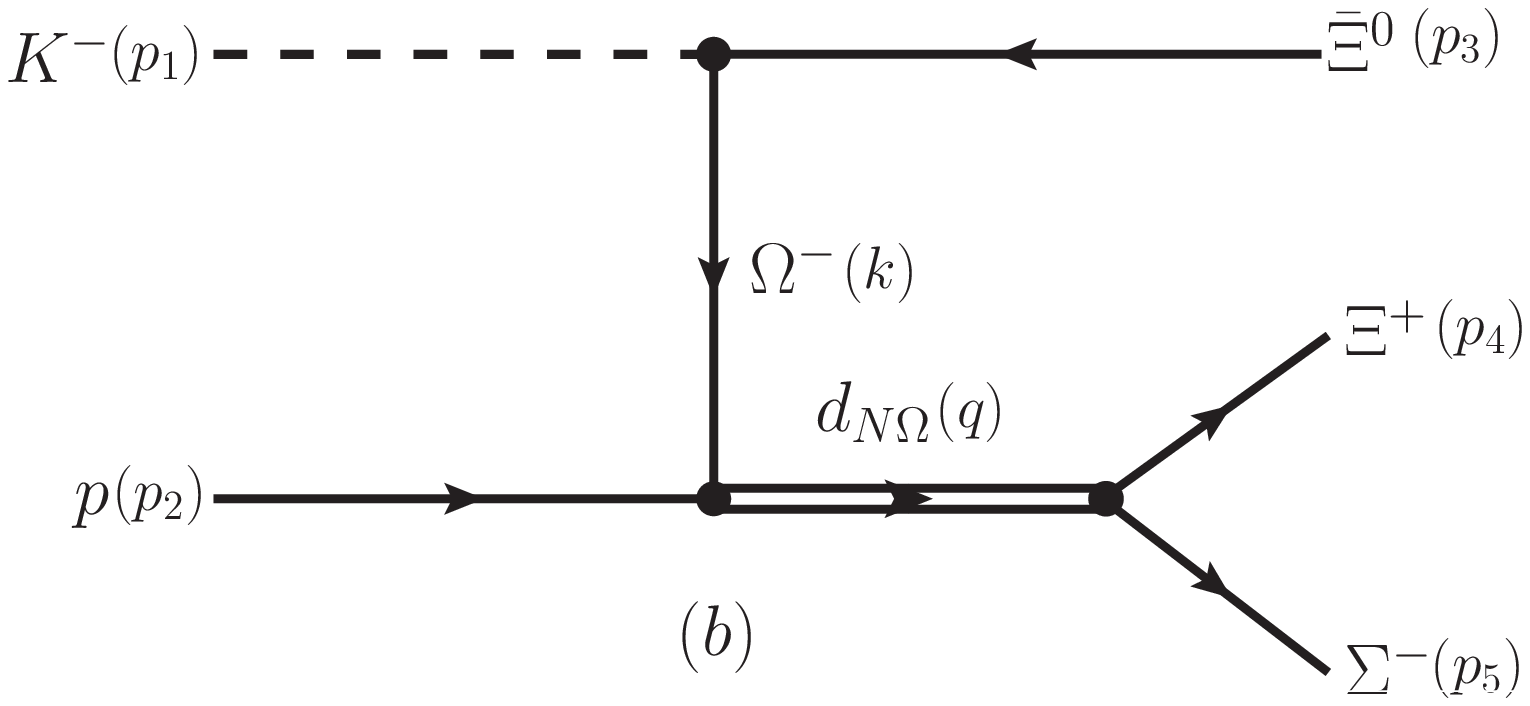}
 \caption{Diagrams contributing to $K^-p \to\Xi^0 \Lambda\bar{\Xi}^0$ (diagram (a)) and
 $K^- p \to
 \Xi^-\Sigma^+ \bar{\Xi}^0$ (diagram (b)).}\label{Fig:Mech2}.
\end{figure*}
\section{Production Processes}
\label{sec:APPROACH}

\subsection{The $K^- p \to d_{N\Omega} \bar{\Xi}^0$ process}

In the present work, we consider the $d_{N\Omega}$ is a dibaryon composed of $N$ and $\Omega$ with the $J(J^P)=\frac{1}{2}( 2^+)$, i.e., $|d_{N\Omega}^0\rangle=|p\Omega\rangle$ and $|d_{N\Omega}^-\langle=|n\Omega\rangle$. The dibayon $d_{N\Omega}$ can be produced in the high energy $K^-p$ interaction process. As indicated in Ref.~\cite{Xiao:2020alj}, the dibaryon $d_{N\Omega}$ could decay into $\Lambda \Xi^0$, which indicate the strong coupling between $d_{N\Omega}$ and $\Lambda \Xi^0$, thus, there exists the $s$ channel contributions to $K^-p\to \bar{\Xi}^0 d_{N\Omega}$ as show in Fig.~\ref{Fig:Mech1}-(a), where the initial $K^- p$ and final $d_{N\Omega} \bar{\Xi}^0$ are connected by $\Lambda$. Moreover, there is the contributions from the $\Omega$ baryon exchange process, which should be the $u$ channel contribution in the strict sense. In the process  $K^-p\to \bar{\Xi}^0 d_{N\Omega}$, there is no $t$ channel digram in the tree level. Empirically, in high energy $K^-p/\pi p$ scattering process, the contributions from the $s$ channels are strongly suppressed~\cite{Liu:2021ojf, Liu:2020ruo, Huang:2015xud, Liu:2008qx}, thus the $d_{N\Omega}$ production in high energy $K^-p$ scattering should occur dominantly by exchanging a $\Omega$ baryon as presented in Fig.~\ref{Fig:Mech1}-(b), while the $s$ channel contribution is ignored.

In the present work, we estimated the cross sections for $K^-p \to d_{N\Omega} \bar{\Xi}$ in an effective Lagrangian approach. The interaction of the $d_{N\Omega}$ dibayon and its components can be described as~\cite{Xiao:2020alj},
\begin{eqnarray}
	\mathcal{L}_{d_{N\Omega} N\Omega}= g_{d_{N\Omega}N\Omega} d^{\mu\nu^\dagger}_{N\Omega} \bar{\Omega}_\mu \gamma_\nu N^c +H.c.,
\end{eqnarray}
where $N^c=C\bar{N}^{T}$, $\bar{N}^c=N^TC$, and $C=i\gamma^2\gamma^0$ is the charge-conjugation matrix, $T$ is the transpose transformation operator. From the above effective Lagrangian, the tensor field of $d_{N\Omega}$ can be constructed by the appropriate combination of a Dirac field for the nucleon and the a Rarita Schwinger field for $\Omega$. The polarization tensor $\epsilon^{\mu\nu}(\vec{p},\lambda)$ could be constructed by the combination of the Dirac field for spin-$1/2$ and a Rarita Schwinger field for spin-$3/2$ i.e,
\begin{eqnarray}
	\epsilon^{\mu \nu}(\vec{p},\lambda) =\sum_{\alpha,\beta} \langle \frac{3}{2} \alpha \frac{1}{2} \beta | 2 \lambda\rangle \psi^\mu_\alpha (\vec{p}) \gamma^\nu \psi_\beta^c (\vec{p})
\end{eqnarray}
with $\lambda=(\pm2, \pm 1,0)$,\ $\alpha=(\pm 3/2, \pm 1/2)$ and $\beta=\pm 1/2$, respectively. The polarization tensor satisfy,
\begin{eqnarray}
	p_\mu \epsilon^{\mu\nu}(\vec{p},\lambda)= p_\nu \epsilon^{\mu\nu}(\vec{p},\lambda)=0,\nonumber\\
\epsilon^{\mu\nu}(\vec{p},\lambda)=\epsilon^{\nu \mu}(\vec{p},\lambda), \ \ \ \epsilon^{\mu}_{\mu}(\vec{p},\lambda)=0\nonumber\\
\epsilon^{\mu \nu \ast} (\vec{p},\lambda) \epsilon_{\mu \nu}(\vec{p},\lambda^\prime)=\delta_{\lambda \lambda^\prime}
\end{eqnarray}

 The effective Lagrangian for $\Omega \Xi K$ interaction reads~\cite{Schutz:1994ue,Ronchen:2012eg,Matsuyama:2006rp,He:2017aps,
Machleidt:1987hj,Liu:2001ce},
\begin{eqnarray}
	\mathcal{L}_{\Omega \Xi K}=
       \frac{g_{\Omega\Xi K}}{m_\pi}\partial_\beta K \bar{\Omega}^\beta \Xi+H.c..
\end{eqnarray}
With the above effective Lagrangians, we can obtain the amplitude corresponding to Fig.~\ref{Fig:Mech1}-(b), which is,
\begin{eqnarray}
\mathcal{M}&=& \frac{g_{\Omega\Xi K}}{m_{\pi}} \left(-ip_{1\beta}\right) \ g_{d_{N\Omega} N\Omega }\ d_{N\Omega}^{\mu\nu} F(k^{2},m^2_{\Omega})
         \nonumber\\
   &\times& \left[ \bar{u}^{c}(p_{2},m_2)\gamma_{\nu}S(k,m_{\Omega})_{\mu\beta}
         \nu(p_{3},m_{3})\right]   \label{Eq:Amp2},
\end{eqnarray}
with $S(k,m_{\Omega})_{\mu\beta}$ to be the propagator of the $\Omega$ baryon, which is,
\begin{eqnarray}
     S(k,m_\Omega)_{\mu\beta}= i\frac{k\!\!\!/+ m_\Omega}{k^2-m_\Omega^2}
    \left[-g_{\mu\beta}
    +\frac{1}{3}\gamma_{\mu\beta}
    +\frac{2k_{\mu}k_{\beta}}{3m_\Omega^{2}}\right.\nonumber \\
   \left. +\frac{\gamma_{\mu}k_{\beta}-\gamma_{\beta}k_{\mu}}{3m_\Omega}\right].
\end{eqnarray}

To depict the internal structure and the off shell effect of the exchanged $\Omega$ baryon, we introduce a form factor $F(k^2, m^2_{\Omega})$ in the amplitude and its concrete form will be discussed later. With the amplitude in Eq. (\ref{Eq:Amp2}), one can obtain the cross section for $K^- p \to d_{N\Omega} \bar{\Xi}^0$ by,
\begin{eqnarray}
\frac{d \sigma}{d \cos \theta}=\frac{1}{32\pi s}\frac{|\vec{p}_f|}{|\vec{p}_i|} \left(\frac{1}{2}
|\overline{\mathcal{M}}|^2\right),
 \end{eqnarray}
where $s=(p_1+p_2)^2$ is the center of mass energy and $\theta$ is the scattering angle, which refers to the angle of outgoing $d_{N\Omega}$ and the kaon beam direction in the center of mass frame. The $\vec{p}_i$ and $\vec{p}_{f}$ are three momentum of the initial kaon beam and the final $d_{N\Omega}$ dibaryon in the center of mass frame, respectively.

\subsection{The $K^- p \to \Lambda \Xi^0 \bar{\Xi}^0$ and $K^- p \to \Sigma^+ \Xi^- \bar{\Xi}^0$ processes}
 As indicated in Ref.~\cite{Xiao:2020alj}, the dibayon $d_{N\Omega}$ dominantly decay into $\Lambda \Xi$ and $\Xi \Sigma$. Thus one can detect $d_{N\Omega}$ in the invariant mass spectrum of $\Lambda \Xi^0$ and $\Xi^- \Sigma^+$ of the processes $K^- p \rightarrow \Xi^0\Lambda \bar{\Xi}^0$ and $K^- p \rightarrow \Xi^-\Sigma^+\bar{\Xi}^0$, respectively. To depict these processes, additional effective Lagrangians related to $d_{N\Omega} \Xi \Lambda$ and $d_{N\Omega} \Xi \Sigma$ are involved. Since $\Xi$, $\Sigma$ and $\Lambda$ have the same $J^P$ quantum numbers, thus these two effective Lagrangians have the same form, which is,

\begin{figure*}[htb]
\includegraphics[width=175mm]{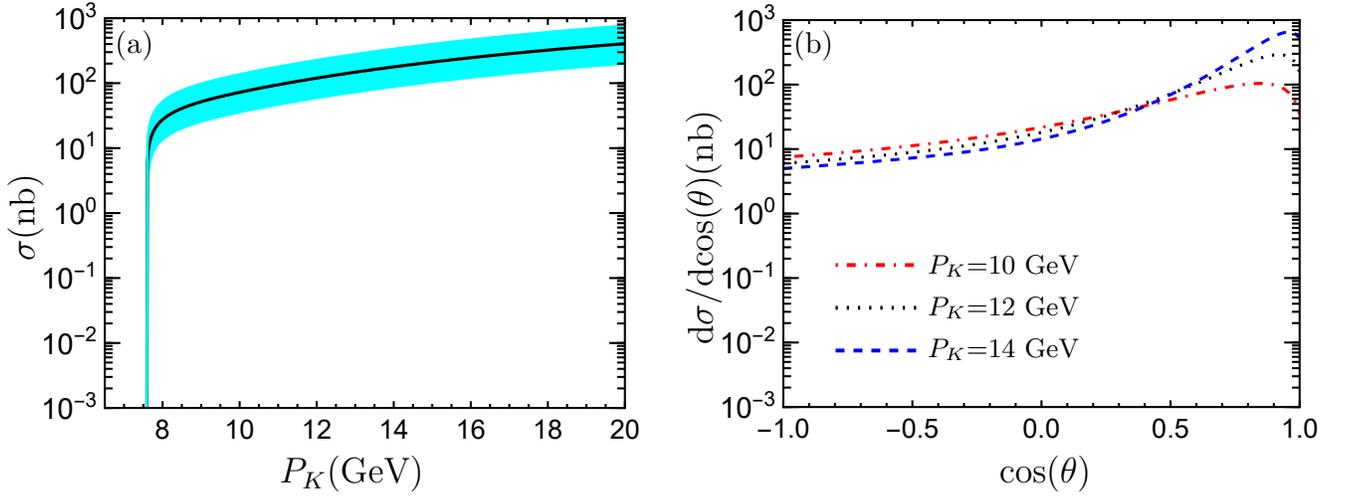}
  \caption{The cross sections for the process $K^-p\to d_{N\Omega} \bar{\Xi}^0$ depending on the beam energy (diagram (a)), and the differential cross sections depending on $\cos(\theta)$ (diagram (b)). }\label{Fig:CS2}
\end{figure*}


\begin{eqnarray}
\mathcal{L}_{d_{N\Omega}Y_1Y_2}&=&i\frac{G_{d_{N\Omega}Y_1Y_2}}{2M_{Y_1}}\bar{Y_1}^c\left(\gamma_{\mu}
           \partial_{\nu}
           +\gamma_{\nu}\partial_{\mu}\right) Y_2 d_{N\Omega}^{\mu \nu} \nonumber\\&&+\frac{F_{d_{N\Omega}Y_1Y_2}}{(2M_{Y_1})^{2}}\partial_{\mu}
           \bar{Y}_1^c\partial_{\nu}Y_2d_{N\Omega}^{\mu \nu}+H.c, \label{Eq:LdYY}
\end{eqnarray}
where $Y_1$, $Y_2$ could be $\Xi$, $\Sigma$ and $\Lambda$. Similar to the case of tensor meson, we can choose $F_{d_{N\Omega} Y_1Y_2}=0$ with tensor dominance hypothesis~\cite{Renner:1970sbf}, and the values of the couplings $G_{d_{N\Omega}Y_1 Y_2}$ will be discussed in next section. With this additional effective Lagrangian, we can obtain the amplitudes corresponding to Fig.~\ref{Fig:Mech2}-(a), which are,
\begin{eqnarray}
 \mathcal{M}_a &=& \Bigg[g_{d_{N\Omega} N\Omega} \bar{u}^c\left(p_2,m_2\right) \gamma^\nu S(k,m_{\Omega})_{\mu \beta} v(p_3,m_3)\Bigg]\nonumber\\ &\times & \left[\frac{g_{\Omega\Xi K}}{m_\pi} (-ip_{1\beta})\right] \mathcal{P}_{d_{N\Omega}}^{\mu\nu \lambda \omega}\Big(q,m_{d_{N\Omega}},\Gamma_{d_{N\Omega}}\Big) \Bigg[i\frac{G_{d_{N\Omega}\Xi\Lambda}}{2m_{\Xi}}
    \nonumber\\
     &\times &  \bar{u}^c(p_5,m_5) \left(\gamma_\lambda (-ip_{4\omega}) +\gamma_\omega (-ip_{4\lambda}) \right) u(p_4,m_4)\Bigg]\nonumber\\
    & \times &
     F\left(k^2,m^2_\Omega\right) F\left(q^2,m^2_{d_{N\Omega}}\right),
             \label{Eq:Amp3}
\end{eqnarray}
where $\mathcal{P}_{d_{N\Omega}}^{\mu\nu \lambda \omega}(q,m_{d_{N\Omega}},\Gamma_{d_{N\Omega}})$ is the propagator of the dibaryon $d_{N\Omega}$, and its concrete form is,
\begin{eqnarray}
&&\mathcal{P}_{d_{N\Omega}}^{\mu\nu\lambda\omega}(q,m_{d_{N\Omega}},\Gamma_{d_{N\Omega}})=\frac{i}{q^2-m_{d_{N\Omega}}^2+im_{d_{N\Omega}}\Gamma_{d_{N\Omega}}}\nonumber\\
&&\quad \quad \quad \quad \quad\times\left[\frac{1}{2}\left(\tilde{g}_{\mu\lambda}\tilde{g}_{\nu\omega}
+\tilde{g}_{\mu\omega}\tilde{g}_{\nu\lambda}\right)
 -\frac{1}{3}\tilde{g}_{\mu\nu}\tilde{g}_{\lambda\omega}\right],  \qquad
 \end{eqnarray}
with $\tilde{g}^{\mu \nu}= -g^{\mu\nu}+q^\mu q^\nu/m^2$. In the above amplitudes, an addition form factor $F(q^2,m^2_{d_{N\Omega}})$ is introduced to depict the internal structure and off shell effects of the $d_{N\Omega}$ and its concrete form will be discussed in the next section. In the same way, one can obtain the amplitude of $K^-p \to \bar{\Xi}^0 \Xi^- \Sigma^+$ corresponding to Fig.~\ref{Fig:Mech2}-(b). With the amplitudes in Eq.~(\ref{Eq:Amp3}), we can obtain the cross sections for the $2\to 3$ processes by,
\begin{eqnarray}
	d\sigma=\frac{1}{8(2\pi)^{4}}
            \frac{1}{\Phi}|\mathcal{M}|^{2}dp_5^0dp_3^0d \cos\theta d\eta,
\end{eqnarray}
where $\Phi=2\sqrt{\lambda(s,m_1^2,m_2^2)}=4|\vec{p}_1|\sqrt{s}$ with $\vec{p}_1$ to be the three momentums of the incident particle $K^-$. $p_3^0$ is the energy of the outgoing $\bar{\Xi}^0$, while $p_5^0$ is the energy of $\Lambda$ in the $K^- p\to \Lambda \Xi^0 \bar{\Xi}^0$ process, and the energy of $\Sigma^+$ in the $K^- p \to \bar{\Xi}^0\Xi^-\Sigma^+ $ process.

\section{NUMERICAL RESULTS and discussions}
\label{sec:NR}

\begin{figure*}
  \includegraphics[width=175mm]{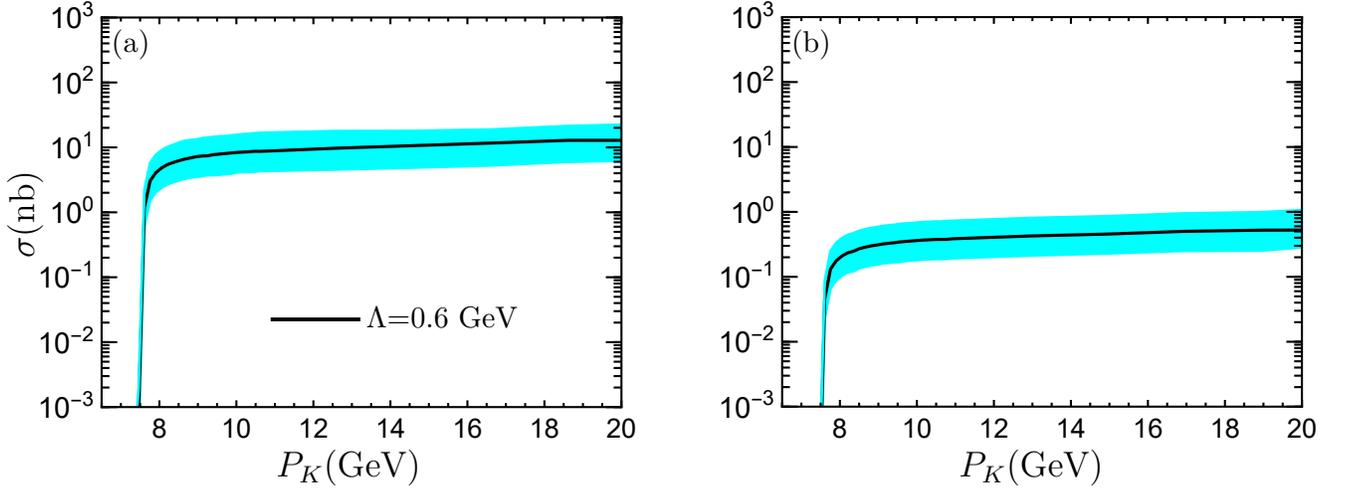}
  \caption{Cross sections for $K^-p \to\Xi^0 \Lambda\bar{\Xi}^0$ (diagram (a)) and $K^- p \to \Xi^-\Sigma^+ \bar{\Xi}^0$ (diagram (b)) depending on the beam energy. }\label{Fig:sici-CS23}
\end{figure*}

\subsection{Form factor and Coupling Constants}
In the present work, we introduce two form factors to depict the internal structures and off-shell effects of the exchanging $\Omega$ baryon and intermediate $d_{N\Omega}$ dibaryon. The specific expression of the factor is~\cite{Chen:2013cpa, Chen:2014ccr, :2019rva, Kim:2015ita},
 \begin{eqnarray}
 F(q^{2},m^2)=\frac{\Lambda^4}{(m^2-q^2)^2+\Lambda^4},\label{Eq:FFs}
 \end{eqnarray}
where $\Lambda$ is model parameter. In principle, the value of the model parameter $\Lambda$ should be determined by comparing the theoretical estimations with the corresponding experimental measurements. However, the  experimental data for the cross sections for the discussed processes are not available at present. In Ref.~\cite{Kim:2015ita}, the authors investigated the cross sections for $\pi^- p \to K^{\ast 0} \Lambda $ and the parameter $\Lambda$ is determined to be 0.55 GeV for the $t$ channel and 0.60 GeV for the $u/s$ channels by comparing the estimated cross sections with the experimental data. With this parameter, they also extended to estimate the cross sections for $\pi^- p \to D^{\ast -} \Lambda_c^+$. In the present work, we take a very similar parameter range, which is $0.55 \ \mathrm{GeV}< \Lambda < 0.65 \ \mathrm{GeV}$, to calculate the cross sections for $K^- p\to d_{N\Omega} \bar{\Xi}^0$ process.

Before the estimations of the cross sections for the discussed processes, the relevant coupling constants should be clarified. As for the coupling constant $g_{d_{N\Omega} N\Omega}$, it can be estimated by the compositeness condition of the composite state. In Ref.~\cite{Xiao:2020alj}, the coupling constant $g_{d_{N\Omega} N\Omega}$ is estimated to be about $1.88\sim 2.38$ with the variation of the model parameter, where the binding energy is set to be $2.46$ MeV. In the present estimation, we take $g_{d_{N\Omega} N\Omega}=1.97$. With this coupling constant, the partial decay widths of $d_{N\Omega} \to \Lambda \Xi^0$ and  $d_{N\Omega} \to \Sigma^+ \Xi^-$ are estimated to be   $582$ and $22.8$ keV~\cite{Xiao:2020alj}, respectively. Together with the effective Lagrangian in Eq.~(\ref{Eq:LdYY}), one can obtain amplitudes of $d_{N\Omega} \to \Lambda \Xi^0$ and $d_{N\Omega} \to \Sigma^+ \Xi^-$, and then the corresponding partial width can be estimated by,
\begin{eqnarray}
\Gamma_{d_{N\Omega} \to ...}=\frac{1}{(2J+1) 8\pi}\frac{|\vec{k_{f}}|}{ M_{d_{N\Omega}}^{2}}|\overline{\mathcal{M}_{d_{N\Omega} \to ...}}|^{2}, \label{Eq:PDW}
 \end{eqnarray}
 where $J=2$ is the angular momentum of $d_{N\Omega}$, $|\vec{k}_f|$ is the three momentum of the daughter particles in $d_{N\Omega}$ rest frame. From Eq.~(\ref{Eq:PDW}) and the partial widths obtained in Ref.~\cite{Xiao:2020alj}, one can estimate the corresponding effective coupling constants, which are $g_{d_{N\Omega} \Lambda \Xi}=7.9\times10^{-2}$ and $g_{d_{N\Omega} \Sigma \Xi}=2.5\times10^{-2}$, respectively. As for the coupling constant $g_{K\Xi\Omega}$, we take $g_{K\Xi\Omega}=2.12$ as in Ref.~\cite{Schutz:1994ue}.

\subsection{Cross sections for $K^- p \to d_{N\Omega} \bar{\Xi}^0$}

With the above preparation, we can evaluate the cross sections for $K^- p \to d_{N\Omega} \bar{\Xi}^0$ depending on the beam energy and the model parameter $\Lambda$. In Fig~~\ref{Fig:CS2}-(a), the solid curve refers to the cross sections for $K^- p \to d_{N\Omega} \bar{\Xi}^0$ with $\Lambda=0.60$ GeV, while the uncertainties of the cross sections are obtained by the variation of the parameter $\Lambda$ from 0.55 GeV to 0.65 GeV. From the figure one can find the cross sections for $K^- p\to d_{N\Omega}\bar{\Xi}^0$ increase sharply near the threshold of $d_{N\Omega} \bar{\Xi}^0$, however, when the beam energy is greater than 9 GeV, the cross sections increase very slowly with the increase of the beam energy. In the considered parameter range, the cross sections are estimated to be $404^{+358}_ {-202}\ \mathrm{nb}$ at $P_K=20$ GeV,  where the center value is estimated with $\Lambda=0.60$, while the uncertainties are resulted from the variation of $\Lambda$ from $0.55$ GeV to 0.65 GeV. In Ref.~\cite{Kim:2015ita}, the cross sections for $\pi p \to D^{\ast -}\Lambda_c^+$ is estimated to be about 13 nb, while the present estimation indicate the cross sections for $K^- p \to d_{N\Omega} \bar{\Xi}^0$ can reach up to 400 nb with the same model parameter, which is about 30 times larger than the one for $\pi p \to D^{\ast -}\Lambda_c^+$~\cite{Kim:2015ita}.

In addition, we also present the differential cross sections depending on $\cos (\theta)$ in Fig.~\ref{Fig:CS2}-(b). We select three typical beam energies as examples, which are $P_K=10,\ 12, \ 14$ GeV with $\Lambda=0.6$ GeV. From the figure one can find that more $d_{N\Omega}$ dibaryon are concentrated in the forward angle area even in the case of $P_K=10 $ GeV, which are resulted from the $\Omega$ exchange.

\subsection{$K^- p\to \Xi^0 {\Lambda} \bar{\Xi}^0$ and $K^- p \to \Xi^- \Sigma^+ \bar{\Xi}^0$}
Besides the $K^- p \to d_{N\Omega} \bar{\Xi}^0$ process, we also estimate the beam energy dependences of the cross sections for $K^- p\to \Xi^0 {\Lambda} \bar{\Xi}^0$ and $K^- p \to \Xi^- \Sigma^+ \bar{\Xi}^0$, where $\Xi^0 \Lambda$ and $\Xi^- \Sigma^+$ are the daughter particles of $d_{N\Omega}$. As indicated in Ref.~\cite{Xiao:2020alj}, the dibaryon $d_{N\Omega}$ dominantly decays into $\Lambda \Xi^0$, and the branching ratio is estimated to be about $95\%$, thus one can experimentally detect $d_{N\Omega}$ in the $\Lambda \Xi^0$ invariant mass distributions of the process $K^-p \to \Xi^0 {\Lambda} \bar{\Xi}^0$ as shown in Fig.~\ref{Fig:Mech2}-(a), where $\Lambda$ can be reconstructed by $p\pi^-$ and $n\pi^0$, while $\Xi^0$ can be reconstructed by the cascade decay processes $\Xi^0 \to \Lambda \pi^0\to p \pi^- \pi^0$ or  $\Xi^0 \to \Lambda \pi^0\to n \pi^0 \pi^0$. Our estimations indicate that the cross sections for $K^-p \to \Xi^0 \Lambda \bar{\Xi}^0$ increase sharply near the threshold and become very weakly dependent on the beam energy. In particular, the the cross section is estimated to be  $13^{+20}_{-7}\ \mathrm{n b}$ at $P_K=20$ GeV, where the center value is estimated with $\Lambda=0.6$, while the uncertainties are resulted from the variation of $\Lambda$ from 0.55 GeV to 0.65 GeV.

Furthermore, the branching ratio of $d_{N\Omega}\to \Xi^- \Sigma^+$ is also sizable, and the final states are charged, which may be easier to be detected. Thus, in the present work, we also estimate the cross sections for $K^- p \to \Xi^- \Sigma^+\bar{\Xi}^0$. As shown in Fig.~\ref{Fig:Mech2}-(b), the beam energy dependences of the cross sections for $K^- p \to \Xi^- \Sigma^+\bar{\Xi}^0$ are very similar to the ones for $K^- p\to \Xi^0 \Lambda \bar{\Xi}^0$ and the magnitude of the cross section is estimated to be  $0.5^{+0.5}_{-0.2}\ \mathrm{n b}$ at $P_K=20$ GeV.

\section{Summary}
\label{Sec:Summary}

The production of the dibaryon $d_{N\Omega}$ is the crucial step of investigating its properties experimentally. The STAR Collaboration at RHIC have detected the dibaryon $d_{N\Omega}$ by measuring the proton-$\Omega$ correlation function in high energy heavy-ion collision. The ALICE Collaboration at LHC also proposed to detect $d_{N\Omega}$ in a very similar processes. By utilizing the secondary kaon beam with the typical momentum to be around 3 GeV, the Hadron experimental Facility at J-PARC proposed to produce $d_{N\Omega}$ by two step reactions, i.e., $K^-p \to \Omega^- K^+ K^{(\ast)0}$, $\Omega d \to \Xi^- \Lambda p$, where the dibaryon $d_{N\Omega}$ is expected to be observed in the $\Xi^- \Lambda$ invariant mass spectrum.

Besides the above two kinds of production processes, we propose to directly produce $d_{N\Omega}$ in the $K^- p \to d_{N\Omega}\bar{\Xi}^0$ process using a secondary kaon beam with the typical momentum to be around 10 GeV in the present work. The cross sections for $K^- p \to d_{N\Omega} \bar{\Xi}^0$ are estimated and we find the  cross section is about several hundreds nanobarn at $P_K=20$ GeV. Moreover, the estimated differential cross sections indicate that the produced $d_{N\Omega}$ are concentrated in the forward angle area.

Considering the fact that $d_{N\Omega}$ dominantly decays into $\Xi \Lambda$, we also estimate the cross sections for $K^- p \to \Xi^0 \Lambda \bar{\Xi}^0$ and $K^- p \to \Xi^+ \Sigma^- \bar{\Xi}^0$, where the $d_{N\Omega}$ dibaryon can be detected in $\Xi^0 \Lambda$ and $\Xi^- \Sigma^+$ invariant mass spectrum, respectively. In particular, the cross sections for $K^- p \to \Xi^0 \Lambda \bar{\Xi}^0$ are estimated to be about ten nanobarn at $P_K=20$ GeV, while the cross section for $K^- p \to \Xi^- \Sigma^+ \bar{\Xi}^0$ is about  20 times smaller than the one for $K^- p \to \Xi^0 \Lambda \bar{\Xi}^0$.

\bigskip
\noindent
\begin{center}
	{\bf ACKNOWLEDGEMENTS}\\
\end{center}
This work is supported by the National Natural Science Foundation of
China under Grants No.11705056, No. 12175037, No.11947224, No.11475192,
No.11975245, and No.U1832173. This work is also supported by the Key
Project of Hunan Provincial Education Department under
Grant No. 21A0039, the State Scholarship Fund of China
Scholarship Council under Grant No. 202006725011, the Sino-German
CRC 110 "Symmetries and the Emergence of Structure in QCD" project by
NSFC under the Grant No. 12070131001, the Key Research Program of
Frontier Sciences, CAS, under the Grant No. Y7292610K1,
and the National Key Research and Development Program of China under
Contracts No. 2020YFA0406300.

\end{document}